\documentclass[aps,preprint,epsfig,rotate]{revtex4}
\usepackage{epsfig}
\usepackage{graphicx}
\usepackage{amsmath}

\newcommand{\beq}{\begin{eqnarray}}
\newcommand{\eeq}{\end{eqnarray}}
\begin{document}

\title{Quantum Control of Ultra-cold Atoms: Uncovering a Novel Connection between
Two Paradigms of Quantum Nonlinear Dynamics}
\author{Jiao Wang$^{1,2}$, Anders S.
Mouritzen$^{3,4}$, and Jiangbin Gong$^{3,5}$\footnote{Corresponding
author: phygj@nus.edu.sg}} \affiliation{$^{1}$Temasek Laboratories,
National University of Singapore, 117542, Singapore
\\$^{2}$Beijing-Hong Kong-Singapore Joint Center for Nonlinear and Complex Systems
(Singapore), National University of Singapore, 117542, Singapore
\\ $^{3}$Department of Physics and Center of Computational Science and Engineering,
National University of Singapore, 117542, Singapore
\\ $^{4}$ Department of Physics and Astronomy, University of Aarhus,
DK-8000, Aarhus C, Denmark
\\ $^{5}$NUS Graduate School for Integrative Sciences and Engineering, Singapore
 117597, Singapore}

\date{\today}

\maketitle
\pagebreak

\begin{center} Abstract\end{center}

\noindent Controlling the translational motion of cold atoms using
optical lattice potentials is of both theoretical and experimental
interest. By designing two on-resonance time sequences of kicking
optical lattice potentials, a novel connection between two paradigms
of nonlinear mapping systems, i.e., the kicked rotor model and the
kicked Harper model, is established.  In particular, it is shown
that Hofstadter's butterfly quasi-energy spectrum in periodically
driven quantum systems may soon be realized experimentally, with the
effective Planck constant tunable by varying the time delay between
two sequences of control fields. Extensions of this study are also
discussed. The results are intended to open up a new generation of
cold-atom experiments of quantum nonlinear dynamics.

%\pacs{05.45.Mt, 05.60.Gg, 03.65.-w, 32.80.Qk}

\vspace{1cm}


\pagebreak
\section{introduction}

One main objective of the field of quantum control
\cite{brumerbook,ricebook} is to use controlled laser-matter
interaction to explore new aspects of quantum dynamics and enhance
our understanding of quantum coherence phenomena. Along this
direction quantum control ideas and techniques are expected to be
very useful for quantum simulation studies, i.e., using controlled
quantum systems to simulate important models. To that end ultracold
atoms and molecules offer promising opportunities due to
experimental advances in Bose-Einstein condensation, the great
controllability of ultracold systems by laser fields, and the long
decoherence time of ultracold systems.

Motivated by our interest in understanding quantum coherence effects
and their control in classically chaotic systems \cite{gongrev}, our
work here focuses on possible cold-atom realizations of fundamental
models of quantum nonlinear dynamics. One important paradigm of
quantum chaos is the kicked rotor model (KRM) \cite{Casati,Izrailev}
whose scaled Hamiltonian can be written as
\begin{eqnarray}
H_{KRM}=\frac{p^{2}}{2}+K \cos(q)\sum_n \delta (t-nT),
\end{eqnarray}
where $q$ ($\in [0,2\pi)$) and $p$ are conjugate coordinate and
momentum variables, and $T$ is the period of the kicking potential.
%spectrum can be discrete or
%continuous depending on the period $T$ of the kicking potential
%therein.
For a general value of $T$, the spectrum of $H_{KRM}$ is discrete
and  the quantum diffusion of the momentum distribution saturates as
a consequence of quantum destructive interference despite the
unlimited classical diffusion. This has been well understood in
terms of a one-dimensional Anderson disorder model \cite{Fishman}.
Interestingly, if the period of the kicking potential $T$ is on
resonance with the quantum recurrence time of the quantum free rotor
dynamics,  i.e., $T$ is a rational multiple of the latter,  then the
associated spectrum will in general consist of continuous bands and
ballistic quantum diffusion emerges. These features of the quantum
KRM have played a significant role in advancing our understanding of
quantum nonlinear dynamics. They have also motivated the
experimental realization of the KRM using cold atoms subject to
kicking optical lattice potentials \cite{Raizen}. Indeed, with the
joint efforts of about ten laboratories worldwide working on the
cold-atom realization of the KRM \cite{KR-exp,KR-exp2,KR-exp3}, many
important dynamical features of the quantum KRM and its variants
have been observed.

Another paradigm of quantum nonlinear dynamics is the kicked Harper
model (KHM) \cite{KH1,KH2,Artuso,prosen,JB-1}.  Using similar
notation as the KRM, the KHM Hamiltonian is given by
\begin{eqnarray} H_{KHM}=(L/T)\cos(p)+K \cos(q)\sum_n \delta (t-nT),
\end{eqnarray} where $L$ and $K$ are two system parameters.  For
later use we note that the classical KHM map is given
by \begin{eqnarray} p_c(n+1)&=&p_c(n)+K\sin[q_c(n)], \nonumber \\
q_c(n+1)&=& q_c(n)-L\sin[p_c(n+1)], \label{cla-KHM}
\end{eqnarray}
where $[q_c(n),p_c(n)]$ denote the values of the classical
coordinate and momentum right before $t=nT$.  The quantum map
associated with each period T is given by
\begin{eqnarray}
U_{KHM}= e^{-i\frac{L\cos(p)}{\hbar}}e^{-i\frac{K\cos(q)}{\hbar}},
\label{KHMmap}
\end{eqnarray}
where $\hbar$ is the effective Planck constant in our scaled unit
system (hence $p=-i\hbar \partial/\partial q$).  For the critical
case of $K=L$, the (quasi-energy) spectrum of the quantum map
$U_{KHM}$ is very similar to the famous Harper model for studies of
two-dimensional electron gases in a strong magnetic field
\cite{hfst}. In particular, the spectrum of $U_{KHM}$ with $K=L$ is
a fractal, often called the ``Hofstadter's butterfly" spectrum
\cite{hfst}. As such, the dynamical properties of the KHM are often
in sharp contrast to those in the KRM. Indeed, complementing the
KRM, the KHM has been regarded as another important paradigm of
quantum chaos, offering a test bed for understanding how a fractal
spectrum is manifested in quantum mapping systems and how the
underlying classical chaos affects the butterfly spectrum. One
important question  thus arises before the research community: How
to experimentally realize or experimentally simulate the KHM?

Two previous studies have proposed to use Fermi-surface electrons in
pulsed fields \cite{fishman2} or a charged particle kicked by a
designed field sequence \cite{dana} to realize the KHM. However,
these proposals, unrelated to ongoing cold-atom experiments of
quantum chaos, have not led to experiments. An intriguing connection
between the KHM and the so-called kicked harmonic oscillator model
\cite{danaprl} might also help realize the KHM, but unfortunately
this connection is subject to the strong restriction of $K=L$. By
contrast, in a recent work \cite{praRapid}, we have briefly reported
that based on existing cold-atom experiments of the KRM, it should
be possible to experimentally realize a quantum version of the KHM
by making use of two sequences of kicking optical lattice potentials
(i.e., making use of a double kicked rotor system, to be explained
later). This unveils for the first time a direct connection between
the two paradigms of quantum nonlinear dynamics, i.e., the KRM and
the KHM. In this paper, we present further discussions and important
details of our finding.  We shall emphasize that the effective
Planck constant of our realization of the quantum KHM can be tuned
by varying the time delay between two sequences of  control fields,
thus offering opportunities for understanding the differences and
similarities between the quantum and classical dynamics. Extensions
of our study to a wider class of periodically driven quantum systems
are also discussed. It is our hope that this contribution can
motivate more interest in quantum control and quantum simulation of
classically chaotic systems.

\section{Kicked Harper model realized by On-resonance double-kicked rotor model}

Our starting point is a modification of the standard KRM, called a
double kicked rotor model (DKRM). That is, within each period $T$,
the free evolution of a rotor is interrupted twice by external
kicking potentials.  Such a model has been experimentally realized
\cite{DKR} and has already attracted considerable interests
\cite{Tania,Tania2}. In terms of the scaled variables used above,
the DKRM Hamiltonian is given by
\begin{eqnarray} H = \frac{p^{2}}{2}+ K_{1} \cos(q)\sum_{n}\delta (t-nT)+
K_{2}\cos(q)\sum_n \delta(t-nT-\eta).
\end{eqnarray}
Clearly, within each period $T$ the rotor experiences two kicks at
$t=nT$ and $t=nT+\eta$, with the two field amplitudes characterized
by $K_1$ and $K_2$. Hence, there are now two sequences of kicking
fields with the same period $T$, and the time delay between the two
sequences of control fields is given by $\eta$. The associated
quantum map $U_{DKRM}$ for a period from $nT+0^{-}$ to
$(n+1)T+{0}^{-}$ is found to be
\begin{eqnarray}
U_{DKRM}=e^{-i(T-\eta)\frac{p^2}{2\hbar}}e^{-i\frac{K_{2}}{\hbar}\cos(q)}
e^{-i\eta\frac{p^2}{2\hbar}}e^{-i\frac{K_{1}}{\hbar}\cos(q)}.
\label{QM}
\end{eqnarray}
Note that for a Hilbert space satisfying the periodic boundary
condition associated with $q\rightarrow q+2\pi$, which should be
the case for a rotor, the momentum eigenvalues can only take
integer values mutiplied by $\hbar$.  Consider now what happens
under the quantum resonance condition, i.e., $T\hbar=4\pi$. Due to
the discreteness of the momentum eigenvalues, one immediately
obtains $e^{-iT\frac{p^2}{2\hbar}}=1$.  Under this resonance
condition, $U_{DKRM}$ is reduced to ${U}^{r}_{DKRM}$,
\begin{eqnarray}
{U}^{r}_{DKRM}&=&
e^{i\eta\frac{p^2}{2\hbar}}e^{-i\frac{K_{2}}{\hbar}\cos(q)}
e^{-i\eta\frac{p^2}{2\hbar}}e^{-i\frac{K_{1}}{\hbar}\cos(q)}
\nonumber \\
&=&
e^{i\frac{\tilde{p}^2}{2\tilde{\hbar}}}e^{-i\frac{\tilde{K}_{2}}{\tilde{\hbar}}\cos(q)}
e^{-i\frac{\tilde{p}^2}{2\tilde{\hbar}}}e^{-i\frac{\tilde{K}_{1}}{\tilde{\hbar}}\cos(q)},
 \label{Ur}
\end{eqnarray}
where we have defined the rescaled momentum $\tilde{p}\equiv \eta p$
and the rescaled kicking amplitudes $\tilde{K}_{1}\equiv \eta
K_{1}$, $\tilde{K}_{2}\equiv \eta K_{2}$. In terms of this rescaled
momentum operator $\tilde{p}$, the effective Planck constant
evidently becomes $\tilde{\hbar}\equiv \eta \hbar$.
%because of the new commutation relation $[q,\tilde{p}]=i\eta{\hbar}$.
Certainly, in a cold-atom realization of this system,  one should
not forget that the cold atoms are actually moving in a flat space
rather than in a compact angular space. Hence, the quantum
resonance condition is relevant only when the initial quantum
state is prepared in
%a definite quasi-momentum state.
a state closely resembling a momentum eigenstate. This is already
well within reach of today's experiments. For example, two recent
experiments \cite{KR-exp2,KR-exp3} realized a KRM on quantum
resonance, using a delocalized Bose-Einstein condensate that
generates appropriate initial states.

Using the rescaled variables defined above and the associated
effective Planck constant $\tilde{\hbar}$,  the quantum map in Eq.
(\ref{Ur}) can now be interpreted in a straightforward manner.
Specifically, within each period $T$, the system is first subject to
one kick, followed by a free evolution interval; then the system is
kicked a second time, followed by a second interval of free
evolution, with the free Hamiltonian for the second free evolution
interval given by $H_{\text{free}}=-\tilde{p}^2/2$. This
interpretation makes it clear that the on-resonance double kicked
rotor model realizes a modified kicked rotor we proposed in Ref.
\cite{JB-3}. Equation (\ref{Ur}) also indicates that the time delay
$\eta$ between the two sequences of the kicking fields offers a
convenient means to vary $\tilde{\hbar}=\eta \hbar$. Hence, the
effective Planck constant $\tilde{\hbar}$ of the quantum map
obtained above can be easily tuned, so long as we keep
$\tilde{K}_{1}$ and $\tilde{K}_2$ constant. This offers a promising
opportunity for studies of the quantum-classical correspondence
associated with the map ${U}^{r}_{DKRM}$.

Consider now a well-defined classical limit of the map $U^r_{RDKR}$,
i.e. the $\eta\rightarrow 0$ limit or equivalently the
$\tilde{\hbar}\rightarrow 0$ limit. Note that this special classical
limit is based on the above on-resonance DKRM, and is hence
unrelated to the conventional classical version of a DKRM (the
conventional classical version of a DKRM is reached by letting
$\hbar\rightarrow 0$ with fixed $K_{1}$, $K_{2}$, $T$, and $\eta$).
Based on the above simple interpretation of ${U}^{r}_{DKRM}$,
%this $\eta\rightarrow 0$ classical limit can be easily obtained.
let us now consider the classical analogs of the quantum observables
$\tilde{p}$ and $q$ right before $t=nT$. These classical quantities
will be denoted $\tilde{p}_c(n)$ and $q_c(n)$, representing the
values of classical canonical variables right before $t=nT$. Then,
after the combined action of the first kick at $t=nT$ and the first
free evolution interval,  $\tilde{p}_c(n)$ and $q_c(n)$ evolve to
$\tilde{p}_c'(n)$ and $q_c'(n)$, with
\begin{eqnarray}
\tilde{p}_c'(n)&=&\tilde {p}_c(n)+ \tilde{K}_1\sin[q_c(n)],
\nonumber
\\ q_c'(n)&=& {q}_c(n)+\tilde {p}_c'(n).
\label{map1}
\end{eqnarray}
Similarly, after the combined action of the second kick and the
second free evolution interval, the total elapsed time is $T$ and
$\tilde{p}_c'(n)$ and $q_c'(n)$ evolve to $\tilde{p}_c(n+1)$ and
$q_c(n+1)$,
\begin{eqnarray} \tilde{p}_c(n+1)&=&\tilde{
p}_c'(n)+ \tilde{K}_2\sin({q}_c'), \nonumber \\
 {q}_c(n+1)&=&
{q}_c'(n)-\tilde {p}_c(n+1).
\end{eqnarray}
The overall classical map in the limit of $\tilde{\hbar}\rightarrow
0$, for the period between $nT+0^{-}$ and $(n+1)T+0^-$, can  be
found by considering the above two steps together, i.e.,
\begin{eqnarray}
\tilde{p}_c(n+1)&=&\tilde{p}_c(n)+\tilde{K}_2
\sin\left[{q}_{c}(n)+\tilde{p}_c(n)+\tilde{K}_1\sin[{q}_c(n)]\right]+
\tilde{K}_1\sin[q_c(n)],
\nonumber \\
q_c(n+1)&=&q_c(n)-\tilde{K}_2
\sin\left[{q}_{c}(n)+\tilde{p}_c(n)+\tilde{K}_1\sin[{q}_c(n)]\right].
\label{map3}
\end{eqnarray}
At first glance this map seems rather complicated, but if we make
a classical canonical transformation
%%%%%%%%%%%%%%%%%%%%%%%%%%%%%%%%%%%%%%%%%%%%%%%%%%%%%%%%%%%%%%%%% Eq 5
\begin{eqnarray}
(q_c,\tilde{p}_c) \rightarrow (Q_c=q_c,
\tilde{P}_c=\tilde{p}_c+q_c), \label{Cc}
\end{eqnarray}
%%%%%%%%%%%%%%%%%%%%%%%%%%%%%%%%%%%%%%%%%%%%%%%%%%%%%%%%%%%%%%%%% Eq 5
then the map of Eq. (\ref{map3}) assumes a much simpler form,
\begin{eqnarray}
\tilde{P}_c(n+1)&=&\tilde{P}_c(n) +\tilde{K}_1 \sin [Q_c(n)],
\nonumber
\\
 Q_c(n+1)&=&Q_c(n) -\tilde{K}_2 \sin [\tilde{P}_c(n+1)]. \label{cmap4}
 \end{eqnarray}
Remarkably,  with the substitutions $\tilde{K}_{1}\rightarrow K$
and $\tilde{K}_{2}\rightarrow L$, the map we obtained in Eq.
(\ref{cmap4}) is
 identical with the classical KHM map in
 Eq. (\ref{cla-KHM})! That
is, for a DKRM under the resonance condition $T\hbar=4\pi$, the
quantum map $U_{DKRM}^{r}$ is simply a quantum version of the KHM,
insofar as its $\tilde{\hbar}\rightarrow 0$ classical limit is
equivalent to a classical kicked Harper model. This surprising
finding exposes a direct connection between the KHM and a kicked
rotor system for the first time.  Because the DKRM was
experimentally realized a few years ago, the finding here indicates
that our quantum version of the KHM can be realized by slightly
modifying the previous experiments \cite{DKR}. Considering the many
fascinating dynamical features of the KHM, we hope that our work
will be able to motivate many new cold-atom experiments of quantum
nonlinear dynamics.

\section{Detailed Results}

%\newpage

\begin{figure}
\vspace{-.5cm}
\includegraphics[width=9.5cm,clip]{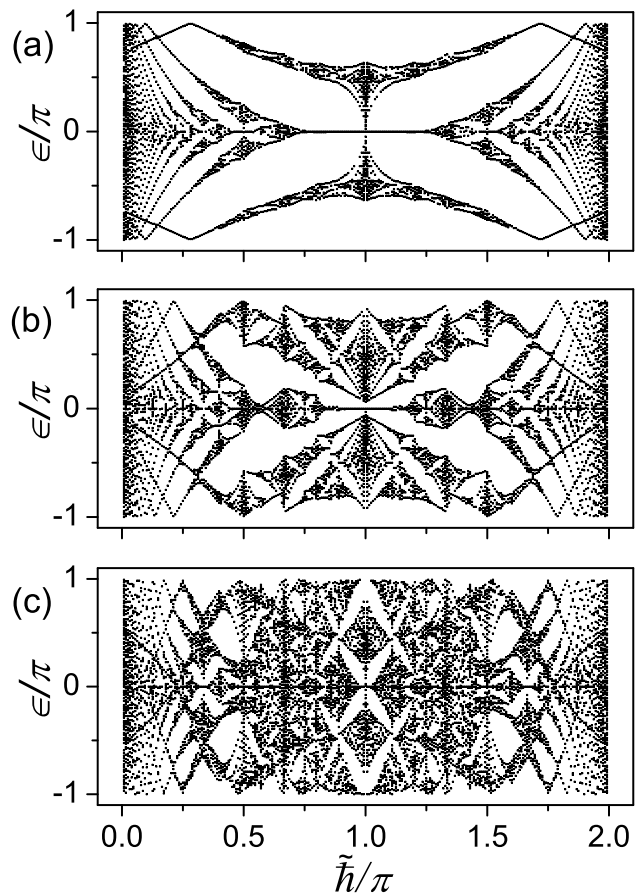}\vspace{-.5cm}
\caption{Quasi-energy spectrum of the cold-atom realization of a
quantum version of the kicked Harper model, achieved by considering
a double kicked rotor model under the main quantum resonance
condition $T\hbar=4\pi$. For panels (a), (b), and (c),  $\tilde
K_1/\tilde \hbar=K_2/\tilde \hbar=2.0$, $3.0$, and $4.0$,
respectively. The Hofstadter's butterfly spectrum seen here is
almost indistinguishable from that calculated from the standard
kicked Harper model.}\label{fig1}
\end{figure}

\subsection{Hofstadter's butterfly spectrum}
In Fig. 1 we show the quasi-energy spectrum of $U_{DKRM}^{r}$ as a
function the effective Planck constant $\tilde{\hbar}$, for three
values of $\tilde K_1/\tilde{\hbar}=\tilde K_2/\tilde{\hbar}$. The
famous Hofstadter's butterfly structure can be clearly seen. Indeed,
we have compared the spectrum of $U_{DKRM}^{r}$ with that of the
$U_{KHM}$ in  Eq. (\ref{KHMmap}), for $\tilde K_1=K=\tilde K_2=L$
and for the same value of the effective Planck constant. The result
is that no difference in the Hofstadter's butterflies can be seen by
the naked eye. For other parameters ($\tilde{K}_1=K\ne
\tilde{K}_2=L$), the remarkable similarity is also observed (not
shown). The fractal property of the spectrum of $U_{DKRM}^{r}$ has
also been checked carefully. For example, for $\tilde K_1=\tilde
K_2=1$, $\tilde \hbar =2\pi/(1+\sigma)$, and
$\sigma=(\sqrt{5}+1)/2$, the generalized fractal dimension
\cite{praRapid} of the spectrum is found to be $D_0\approx 0.5$.
This is identical to that of the KHM, with $K=L=1$ and the same
value of the effective Planck constant.

The results in Fig. 1 bring up an interesting question. Is the
butterfly spectrum of $U_{DKRM}^{r}$ mathematically the same as that
of the KHM?  That is, does there always exist a unitary
transformation to transform $U_{KHM}$ in Eq. (\ref{KHMmap}) to
$U_{DKRM}^{r}$? The answer is no.  Even though a canonical
transformation of the classical limit of $U_{DKRM}^{r}$ is
equivalent to the classical limit of $U_{KHM}$, their quantum
spectra can still be different due to the periodic boundary
condition associated with their Hilbert spaces. Qualitatively, this
is because a classical canonical transformation does not necessarily
have a unitary transformation analog in the quantum case.  In other
words,
%aspects of a particular quantization
a particular quantization strategy can induce differences between
the butterfly spectra of $U_{DKRM}^{r}$ and that of $U_{KHM}$.

Especially, we have shown that for fixed $K/\hbar$ and $L/\hbar$,
the spectrum of $U_{KHM}$ has a period $2\pi$ in $\hbar$, and is
reflection symmetric about $\hbar=\pi$. By contrast, the spectrum of
$U^{r}_{DKRM}$ (for fixed $\tilde K_1/\tilde \hbar$ and $\tilde
K_2/\tilde \hbar$ ) has a period $4\pi$ in $\tilde \hbar$, and is
reflection symmetric about $\tilde \hbar=2\pi$ instead. Also
significant, it can be proved that the spectrum of $U^{r}_{DKRM}$ is
invariant if we swap $\tilde K_1$ and $\tilde K_2$.

As an example of spectral differences, we find that so long as
$\tilde K_1\ne \tilde K_2$, the spectrum of $U^{r}_{DKRM}$ is in
general continuous. This is markedly different from the KHM, where
the spectrum can change from being continuous to being discrete if
we swap the values of $K$ and $L$ \cite{Artuso,prosen}. Other subtle
spectral differences between $U^{r}_{DKRM}$ and $U_{KHM}$,
especially those regarding the butterfly spectrum's sub-band widths
in cases of $\tilde {\hbar}=\hbar=2\pi r/s$ ($r$ and $s$ being
integers), have also been studied \cite{JB-4} but will not be
discussed here.

\subsection{Quantum Diffusion Dynamics}
Our cold-atom proposal for realizing one quantum version of the KHM
has presented exciting opportunities for experimental studies of
quantum diffusion dynamics with a butterfly spectrum.  Though the
fractal butterfly spectrum seen above is not a direct experimental
observable, the
%statistical
fractal properties of the spectrum do govern the quantum dynamics
\cite{Geisel}.  A number of previous studies have contributed to
understanding the manifestations of a fractal spectrum in the
quantum dynamics.

\begin{figure}
\vspace{-0.25cm}
\includegraphics[width=12cm,clip]{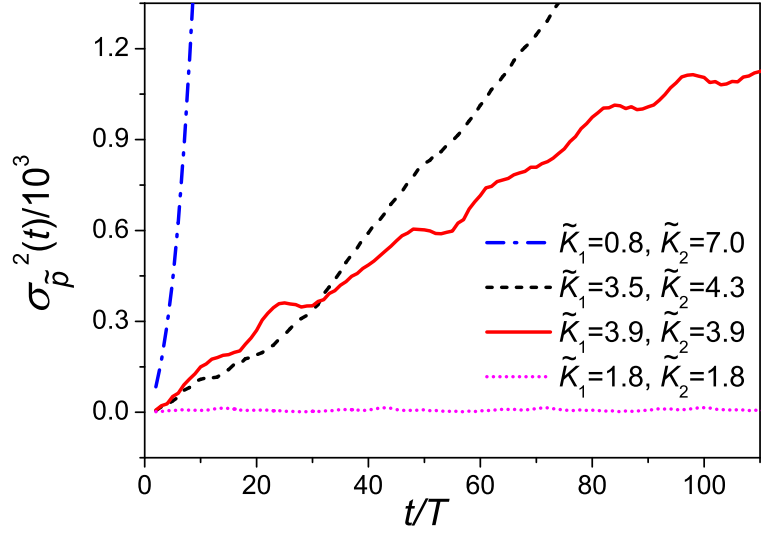}
\caption{Time dependence of the momentum variance
$\sigma_{\tilde{p}}^{2}(t)$ in the cold-atom realization of the
kicked Harper model. The main features range from localization,
subdiffusion to almost ballistic diffusion. In the dotted line case
$\tilde \hbar=6\pi/19$;  for other three cases $\tilde \hbar=1$. The
initial state is taken as a zero momentum state, which can be
approximately realized by use of a dilute Bose-Einstein condensate
with very large coherence length.} \label{fig2}
\end{figure}

To motivate experimental work based on our finding we present here
some remarkable numerical results of the quantum diffusion dynamics
associated with $U_{DKRM}^{r}$.
%In particular,
We choose to examine the time dependence of the momentum variance of
a time-evolving wavepacket $|\varphi(t)\rangle$. This variance is
given by $\sigma_{\tilde{p}}^2(t)\equiv \tilde{\hbar}^2\sum_l
|\langle \varphi(t)|l\rangle|^2 (l-l_0)^2$, where $\{|l\rangle\}$
are the eigenstates of $\tilde{p}$ with the eignevalue
$l\tilde{\hbar}$, $|l_0\rangle$ is the initial state, and $l_{0}$ is
the time-independent mean value of $\tilde{p}/\tilde{\hbar}$.  Note
that $\sigma_{\tilde{p}}^2(t)$ is a quantity easily measurable in
current cold-atom experiments of the KRM.  In our numerical
calculations we assume $l_0=0$ but we have checked that the results
presented below
do not depend on the initial condition we choose. % and that $\langle
%\tilde p(t)\rangle\equiv \tilde{\hbar}\sum_l |\langle
%\varphi(t)|l\rangle|^2 l=\tilde \hbar l_0$.

First of all, for both $\tilde K_1\ll \tilde K_2$ and $\tilde K_1\gg
\tilde K_2$, the dynamics of $U_{DKRM}^{r}$ typically displays
almost ballistic diffusion with $\sigma_{\tilde{p}}^2(t)\sim t^2$.
One such example is represented by the dot-dashed line in Fig. 2. It
is also computationally found that if we swap $\tilde K_1$ and
$\tilde K_2$, then the time dependence of $\sigma_{\tilde{p}}^2(t)$
does not change. This feature is consistent with our previous
discussion and does not exist in the dynamics of $U_{KHM}$.

Besides, for $\tilde{K}_1=\tilde{K}_2$ and for a generic value of
$\tilde{\hbar}$, the quantum diffusion dynamics is expected to be
drastically different because the spectrum is a fractal (see Fig.
1). In Fig. 2 we show one such computational example with $\tilde
K_1=\tilde K_2=3.9$ and $\tilde \hbar=1$ (solid line). It is seen
that the quantum diffusion dynamics in this case is much slower than
the previous ballistic diffusion case.  To further understand this
case we follow the dynamics over a time scale of $t=10^4 T$ and then
fit $\sigma_{\tilde{p}}^2(t)$ by a power law. This fitting yields
$\sigma_{\tilde{p}}^2(t)\sim t^\alpha$ with the quantum diffusion
exponent $\alpha\approx 0.82$, indicating an anomalous diffusion.
Applying an existing theory \cite{Artuso2} that relates the quantum
diffusion exponent $\alpha$ to the fractal dimension of the
spectrum,  we infer that in the case of $\alpha\approx 0.82$, the
Hausdorff-dimension of the fractal spectrum is $D_H\approx 0.41$.
Interestingly, if we slightly mismatch $\tilde K_1$ and $\tilde K_2$
(dashed line in Fig. 2), then it is found that after a transient
time of about $10^2T$, $\sigma_{\tilde{p}}^2(t)$ will display
ballistic diffusion
%again.
like in the first example. As such, by merely slightly tuning the
relative strengths of the two sequences of the control fields
without even changing their average strength
($\tilde{K}_1+\tilde{K}_2)/2$, one may
%already
generate many
remarkable and qualitatively different features in the quantum
diffusion dynamics.

Let us finally discuss the result of the dotted line in Fig. 2. In
this case, $\tilde \hbar=6\pi/19$, $\tilde{K}_1=\tilde{K}_2=1.8$,
and $\sigma_{\tilde{p}}^2(t)$ is seen to saturate at very small
values. Computationally, this saturation behavior is found to
persist for times larger than $10^{6}T$. Hence the quantum
diffusion dynamics in this case constitutes an example of strong
localization. This localization behavior is related to the fact
that the spectrum sub-band width of the butterfly of
$U_{DKRM}^{r}$ may become vanishingly small as
$\tilde{K}_1=\tilde{K}_2$ decreases \cite{JB-4}. We believe that
the localization observed here represents a novel type of
dynamical localization in quantum nonlinear dynamics. Our ongoing
theoretical work will soon offer more insights into this
\cite{JB-4}. Experimentally speaking, observing such kind of
localization behavior would be a strong indication that our
cold-atom version of the KHM has been cleanly realized.

\subsection{Extension to Double Kicked Rotor Systems on Higher-Order Resonances}

%Hence we can roughly divide the parameter space
%spanned by $\tilde K_1$ and $\tilde K_2$ into $\sim t^2$ fast
%diffusion region of $\tilde K_1\ne\tilde K_2$ and that of $\tilde
%K_1=\tilde K_2$ associated with the fractal butterflies.

\begin{figure} \vspace{-.5cm}
\includegraphics[width=9.5cm,clip]{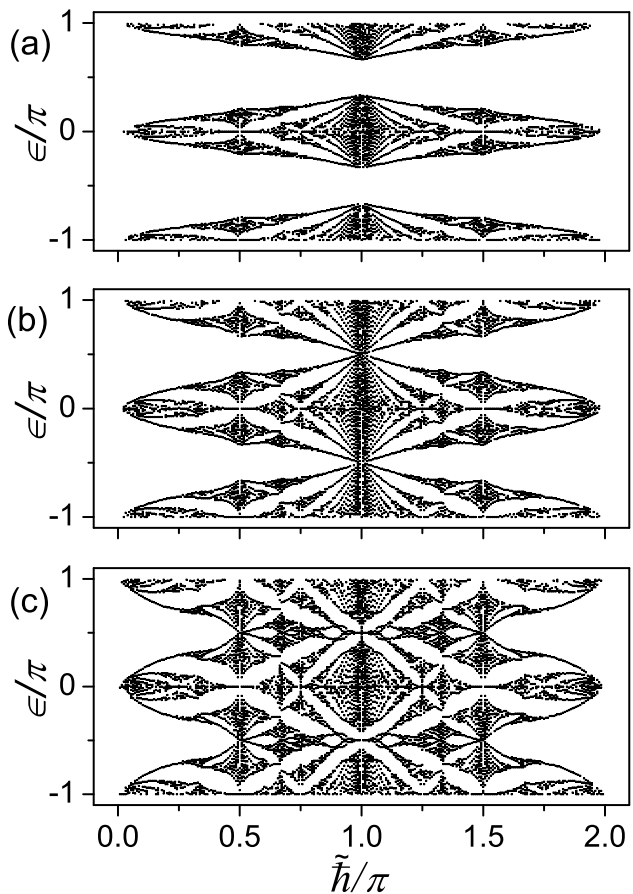}\vspace{-.5cm}
\caption{Quasi-energy spectrum of a double kicked rotor system under
the quantum anti-resonance condition $T\hbar=2\pi$.  In panels (a),
(b) and (c), $\tilde K_1/\tilde \hbar=\tilde K_2/\tilde \hbar=1.2$,
$\pi/2$, and $2.0$, respectively. In (b), three main branches of the
spectrum pattern begin to touch each other. The complex spectrum
pattern as a function of $\tilde{\hbar}$ is called ``generalized
Hofstadter's butterfly" in the text.} \label{Fig3}
\end{figure}

So far, we have
%shown
shown that by considering a double kicked rotor under the resonance
condition $T\hbar=4\pi$, a cold-atom realization of the KHM is
within reach of today's experiments.  It now becomes interesting to
ask whether Hofstadter's butterfly spectrum also exists for a double
kicked rotor system under higher order quantum resonances, i.e.,
under the condition $T\hbar=4\pi\nu/\mu$, with $\nu$ and $\mu$ being
integers and with $\mu>1$. Our preliminary study suggests that this
is the case, thus opening up studies of quantum diffusion dynamics
in a wider class of quantum mapping systems.

As an example, we present in Fig. 3 the spectrum of the DKRM at the
so-called quantum anti-resonance, with $\nu=1$, $\mu=2$.   For
$\tilde K_1/\tilde \hbar=\tilde K_2/\tilde{\hbar}=0$,  the quantum
map operator $U_{DKRM}$ in Eq. (\ref{QM}) reduces to $U=e^{-i\pi
p^2/\hbar^2}$, yielding only two nonequivalent values of the
quasi-energy, i.e.,  $0$ and $\pm\pi$. As $\tilde K_1/\tilde
\hbar=\tilde K_2/\tilde{\hbar}$ increases,  the complex structure of
the spectrum ``grows'' around the points 0 and $\pm\pi$. Due to this
growth, the spectrum for the case of $K_1/\tilde \hbar=\tilde
K_2/\tilde{\hbar}=1.2$ is already quite complicated, as seen in Fig.
3(a).  When $\tilde K_1/\tilde \hbar=\tilde K_2/\tilde{\hbar}$
reaches $\pi/2$ [see Fig. 3(b)], three main branches of the spectrum
begin to touch one another, yielding an overall pattern that can be
regarded as a generalized Hofstadter's butterfly spectrum.  With the
values of $\tilde K_1/\tilde \hbar=\tilde K_2/\tilde{\hbar}$
increased even further, the generalized Hofstadter's  ``butterfly"
pattern gets larger and possesses more fine structures.  Unaware of
any study with similar results, we think that the generalized
``butterfly" spectrum found here and those associated with other
higher-order quantum resonances will stimulate considerable
theoretical work.
 The results further strengthen
the view that using controlled laser-matter interactions, novel
quantum dynamics models may be generated and explored.

\section{Conclusion}
To conclude,  we have shown that by considering a  cold-atom
realization of an on-resonance double kicked rotor model we can
realize a quantum version of the kicked Harper model from a kicked
rotor system.  To the naked eye, the butterfly spectrum we obtain
from our version of a kicked Harper model is almost
indistinguishable from that of the standard kicked Harper model.  We
have also stressed that the effective Planck constant of the quantum
kicked Harper model realized here can be easily tuned by varying the
time delay between two sequences of control fields. Extending our
considerations to double kicked rotor systems under higher-order
quantum resonances, we have shown that an entirely new class of
quantum mapping systems with generalized Hofstadter's butterfly
spectra can be studied, both theoretically and experimentally. These
results
%will lead to
present many new opportunities in the studies of quantum nonlinear
dynamics in external control fields. Indeed, with controlled
interactions between laser fields and cold atoms, there are still
so much to see and explore.

\section{Acknowledgments}
Two of the authors (J.W. and J.G.) are very grateful to Prof. C.-H.
Lai for his support and encouragement. One of the authors (J.W.)
acknowledges support from Defence Science and Technology Agency
(DSTA) of Singapore under agreement of POD0613356.  One of the
authors (A.S.M.) is supported by a Villum Kann Rasmussen grant. One
of the authors (J.G.) is supported by the start-up fund (WBS No.
R-144-050-193-101/133), and the NUS ``YIA'' fund (WBS No.
R-144-000-195-123) from the office of Deputy President (Research \&
Technology), National University of Singapore.

%\newpage
%\begin{center}J. Wang et al, Fig. 1\end{center}

%\begin{figure}
%\vspace{-.5cm}
%\includegraphics[width=9.5cm,clip]{figure1.EPS}\vspace{-.5cm}
%\includegraphics[width=9.5cm,clip]{figure1-low.eps}\vspace{-.5cm}
%\caption{J. Wang et al, Fig. 1}\label{fig1}
%\end{figure}

%\pagebreak

%\begin{center}J. Wang et al, Fig. 2\end{center}

%\begin{figure}
%\vspace{-0.25cm}
%\includegraphics[width=12cm,clip]{figure2.EPS}
%\caption{J. Wang et al, Fig. 2} \label{fig2}
%\end{figure}

%\pagebreak

%\begin{center}J. Wang et al, Fig. 3\end{center}
%\begin{figure} \vspace{-.5cm}
%\includegraphics[width=9.5cm,clip]{figure3.EPS}\vspace{-.5cm}
%\includegraphics[width=9.5cm,clip]{figure3-low.eps}\vspace{-.5cm}
%\caption{J. Wang et al, Fig. 3} \label{Fig3}
%\end{figure}

\end{document}